# Search Sources of Cosmic Rays Ultrahigh Energy

A.A. Mikhailov

Yu.G. Shafer Institute of Cosmophysical Research and Aeronomy SB AN,
Lenin ave. 31, Yakutsk, Russia, mikhailov@ikfia.ysn.ru.

## Abstract

The arrival directions of ultrahigh energy extensive air showers (EAS) by Yakutsk, AGASA and P. Auger data are considered. It is found that the arrival directions of EAS with a deficit muons by Yakutsk data are not isotropy. Majority of these EAS form doublets which have maximum at side anticenter of Galaxy. It is shown that some EAS by data of Yakutsk, AGASA, P.Auger correlate with pulsars. These pulsars are situated near Input and Output Local arm Galaxy Orion. The majority of these pulsars have a short period rotate around of their axes. The problem of cosmic ray origin is discussed.

At first we have analyzed extensive EAS by data of Yakutsk EAS array. We consider showers with energy $E>5.10^{18}$ eV, with zenith angles $\vartheta<60°$ and the axes lying inside of perimeter of array. The accuracy of definition of solid angles of arrival directions is 5-7°, energy - 30%.

Earlier we have been found showers with deficit content muons [1]. Theoretical calculations show [2] that the content muons of showers reflect a mass composition of the particles which have formed them. Probably, EAS with the usual content muons are formed by the charged particles, EAS with the deficit content muons - neutral particles. Among EAS with the deficit content muons we found some showers without muon component (a threshold of registration muons by detectors is > 1 GeV). If a probability of registration P showers without muons was $P>10^{-3}$ the given EAS was excluded from consideration. We have found 22 EAS without muons and 5 EAS with poor muons - a density muons at distance>100 m from axis was less, than it is expected more than $3\sigma$.

In Fig.1 on the equal - exposition map of celestial sphere the distribution of these EAS is shown. Distribution of EAS with deficit muons is not isotropic, from the side of a galactic plane some excess of an observed number of EAS is observed: $n(|b|<30°)/n(|b|>30°) = 1.9 \pm 0.7$ (b – a galactic latitude) instead in a case of isotropy 1.2 according to [3].

We have found among these EAS 5 doublets and from them 4 doublets are located at one region of a celestial sphere: $\delta =20-75°$ and $60°<RA<80°$. The fifth doublet which consists of two EAS: one without muons and the other – with poor muons, is located near Input of the Local Arm of the Galaxy Orion.

We are interested in this maximum of doublets and consider a distribution of usual EAS on a right ascension RA. We divided a observed region of energy $>10^{18}$ eV into 4 intervals: 1) $10^{18}$ - $5.10^{18}$ eV, 2) $5.10^{18}$ - $10^{19}$ eV, 3) $10^{19}$ - $4.10^{19}$ eV, 4) $> 4.10^{19}$ eV and the distribution of EAS on the right ascension was analyzed by harmonic functions of Fourier.

In Fig.2 amplitudes and phases of 1-st harmonic are shown. In interval energy $10^{19}$ - $4.10^{19}$ eV the amplitude of 1-st harmonic is close to statistically-significant value: $A_1=18.8 \pm 6.3\%$ and its phase RA=14°; probability of chance of amplitude is equal P ~ $1.2 \times 10^{-2}$. Number of EAS is equal 502. The amplitude of 2-nd harmonic is - 5 %. Such value amplitude of 1-st harmonic by Yakutsk data we received earlier [5], but here the statistics of data has increased. Note, that phase of 1-st harmonic at $E \sim 10^{18}$ eV from the Local Arm of Galaxy at RA~300° varies gradually with energy to RA ~ 90° at $E \sim 4.10^{19}$ eV where 4 doublets are located.

Further we consider distribution EAS on the right ascension also. We have divided a region of energy into 3 intervals: 1) $5.10^{18}$ - $10^{19}$ eV, 2) $10^{19}$ - $4.10^{19}$ eV, 3)$> 4.10^{19}$ eV.

In Fig.3 distribution of particles is shown. We observe a maxima of the distribution particles at 2 first intervals of energy at coordinates 60°<RA<90°. From this region of celestial sphere we observe neutral and charged particles. This fact is difficult to explain by an extragalactic origin of cosmic rays because extragalactic particles are expected to be more isotropy (in our case arrival directions of particles with different composition are not isotropy).

We have found out 27 EAS with deficit content muons (Fig.3). From them only 17 EAS are inside of angular distance 6° from pulsars (we choose 6° because at this angular distance from pulsars a correlation between arrival direction of EAS with usual muons and pulsars at $E>8.10^{18}$ eV was maximum [6]). Arrival directions of 17 EAS are marked by red circles. Distribution of these EAS which correlate with pulsars is isotropic in the main, but the majority of them are observed near a galactic plane.

Further we have considered the arrival directions EAS with energy $E>4.10^{19}$ eV and with the usual content muons. Do they correlate with pulsars?. Thus, we have selected pulsars which are situated at angular distances <6° from arrival directions of EAS. According by Yakutsk data we have found such 19 EAS from 34 (these EAS are noted by red colors, Fig.4), according to array AGASA - 21 EAS from 57 (Fig.5), according to array P. Auger - 10 EAS from 27 (Fig.6). Arrival directions of these particles, which are correlated with pulsars, are situated near a galactic plane and at Input (Yakutsk and AGASA) and at Output (P. Auger) of the Local arm of the Galaxy Orion. Correlation between arrival directions EAS and pulsars from this direction is not possible to explain by extragalactic origin of particles. Note, earlier we found anisotropy of arrival directions particles with energy $E>4.10^{19}$ eV from side Input and Output of the Local arm by data of these arrays [7].

We consider the rotation periods of chosen pulsars. Ratio number of pulsars with periods $P_0 <P$ to the number of pulsars which have the periods $P_0>P$ is shown in Fig.7 (in case of Yakutsk array we have considered EAS with a deficit and usual muons). As seen from Fig.7 majority of pulsars have short periods $P_0$ than it is expected according to the catalogue of pulsars. Some authors showed that short period pulsars can accelerate heavy nuclei up to $10^{20}$ eV [8,9].

## Summary

The analysis of EAS with a deficit content muons by Yakutsk data show that third part of them form doublets which are located in mainly on right ascension 60°<RA<90°. At this coordinates maximum distribution of usual EAS at energy $E>5.10^{18}$ eV by Yakutsk data is observed.

It is found that particles with energy $E>4.10^{19}$ eV by data of Yakutsk, AGASA and P.Auger correlate some pulsars which are situated near Input and Output of Local arm of Galaxy Orion. Majority of these pulsars have a short period rotation around their axes. Most likely cosmic rays have a galactic origin.

This paper has been supported by RFBR (project N 08-02-0497)

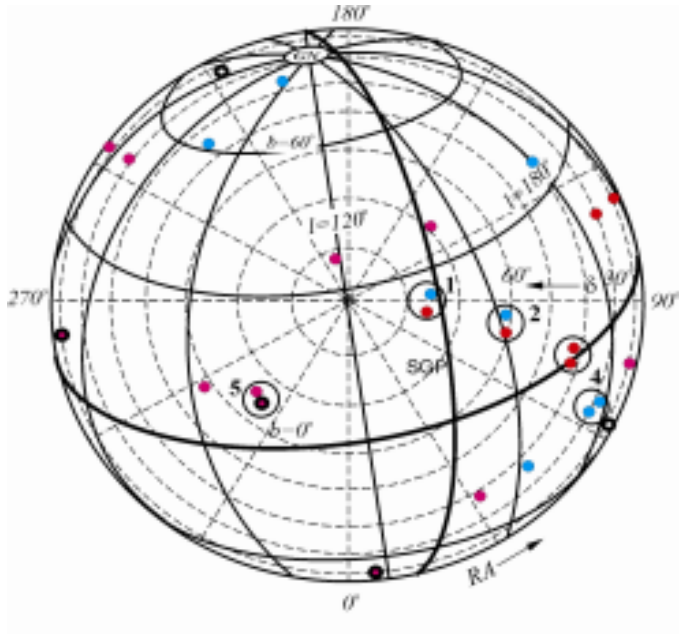

Fig.1. Distribution of EAS with deficit muons in a celestial sphere: •,• - EAS without muons, o – EAS with poor muons. δ - declination, RA – right ascension, b – galactic latitude, l – longitude. Red circles – EAS correlated with pulsars.

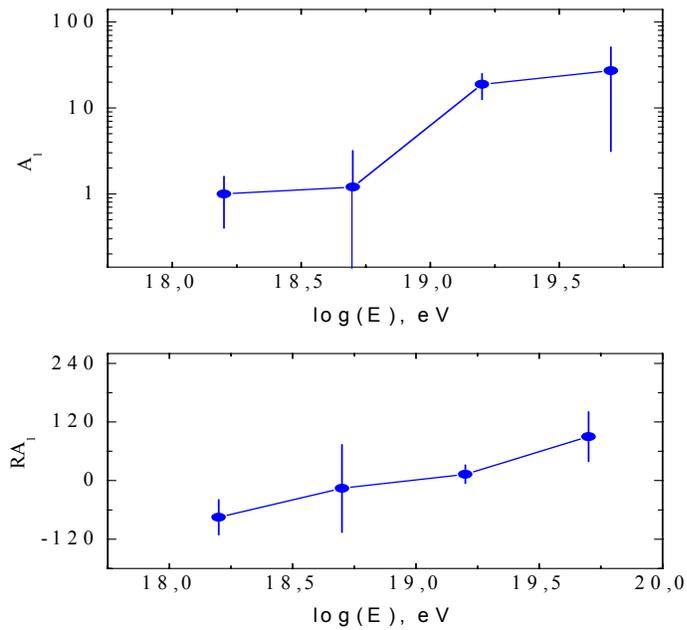

Fig.2. Amplitudes $A_1$ and phase's $RA_1$ of the 1-st harmonic are shown in energy intervals.

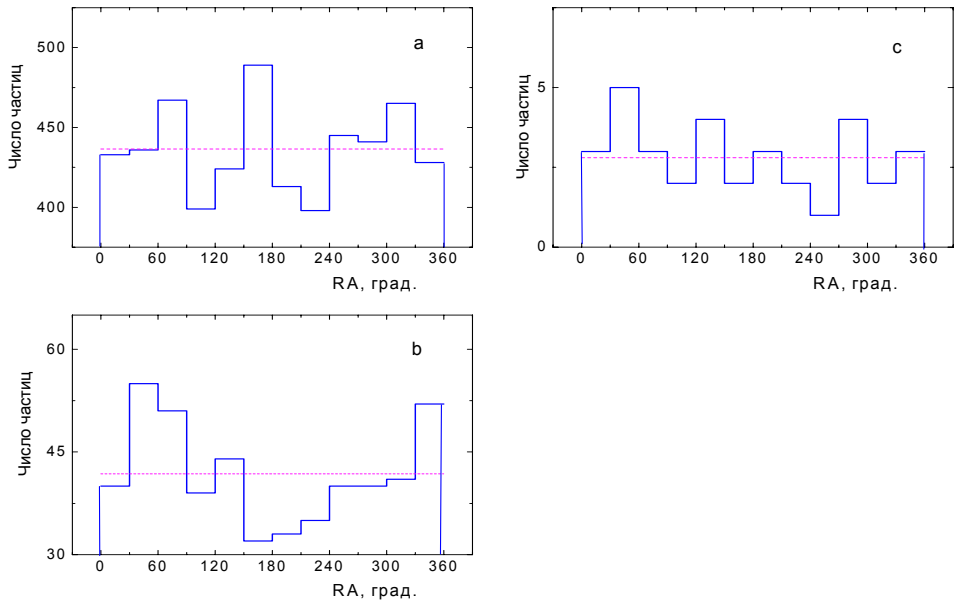

Fig.3. Distribution particles in energy intervals: a) $E=5 \cdot 10^{18} - 10^{19}$ eV; b) $10^{19} - 4 \cdot 10^{19}$ eV; c) $E > 4 \cdot 10^{19}$ eV, red lines – an expected number of particles in a case of isotropy.

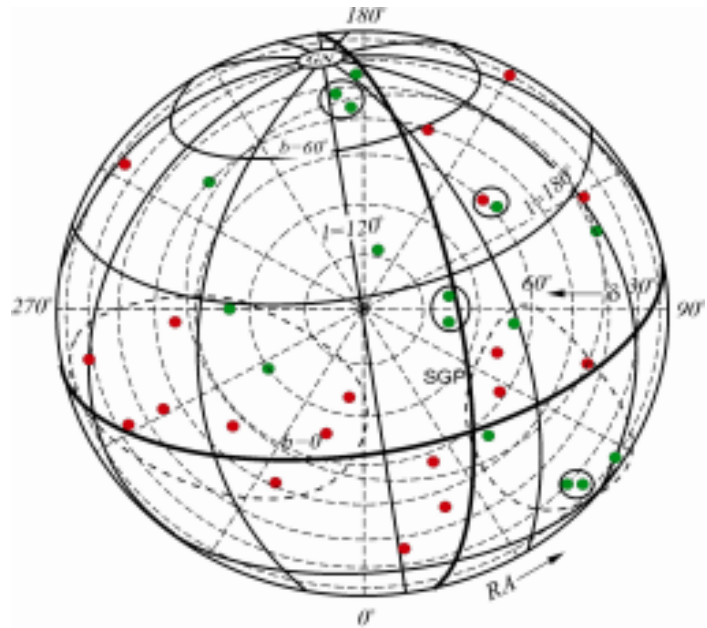

Fig.4. Yakutsk: a distribution particles with $E > 4 \cdot 10^{19}$ eV. Red circles – EAS correlated with pulsars.

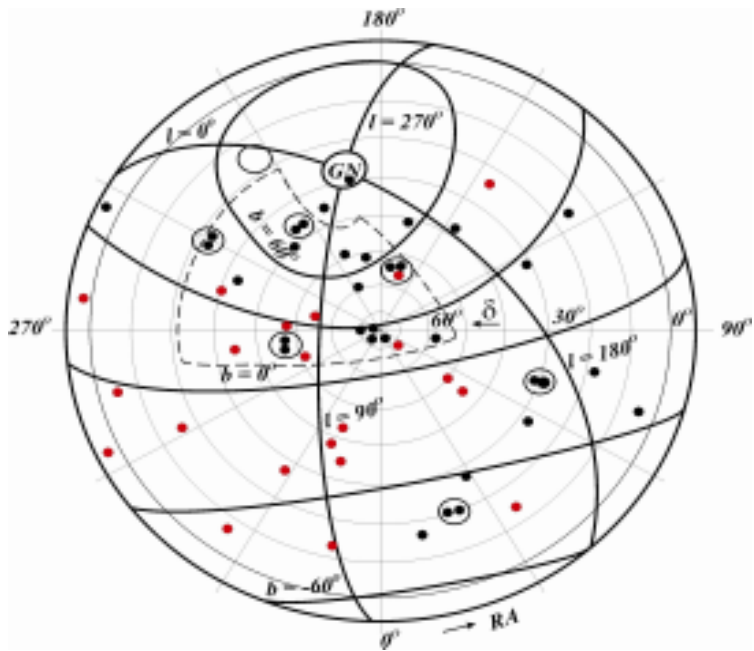

Fig.5. AGASA: distribution particles with E>4.10$^{19}$ eV. Red circles – EAS correlated with pulsars.

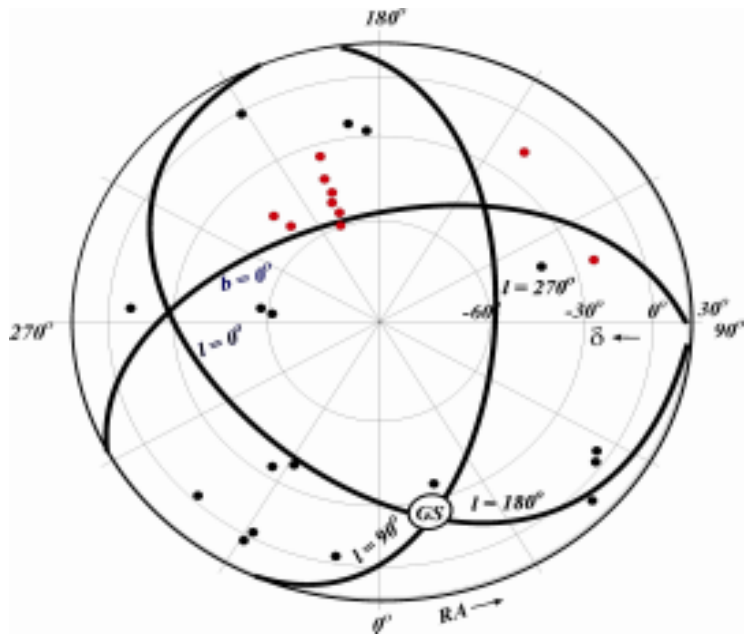

Fig.6. P. Auger: distribution particles with E>4.10$^{19}$ eV. Red circles - EAS correlated with pulsars.

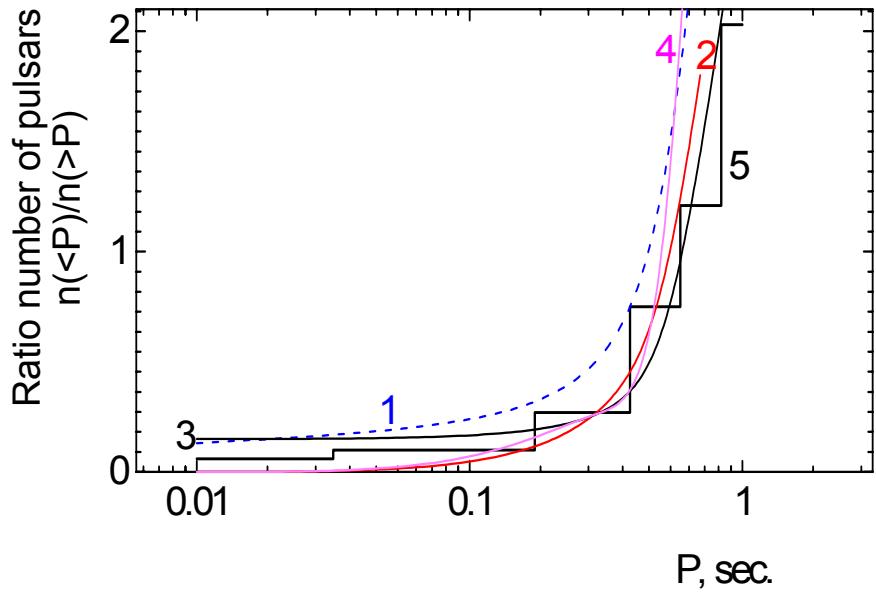

Fig.7  Ratio number of pulsars with period $P_0$ - $n(P_0<P)/n(P_0>P)$:

1 - pulsars correlated with EAS of deficit muons, Yakutsk;

2 - pulsars correlated with usual EAS, Yakutsk;

3 - pulsars correlated with EAS, AGASA;

4 - pulsars correlated with EAS, P. Auger;

5 - pulsars according to the catalogue [5].